# Super-resolution measurement of thermo-optic coefficient of KTP crystal based on phase amplification

**WUZHEN LI,[1,2,3] ZHIYUAN ZHOU,[1,2,3,*] GUANGCAN GUO, AND BAOSEN SHI[1,2,3,†]**

[1]*CAS Key Laboratory of Quantum Information, University of Science and Technology of China, Hefei, Anhui 230026, China*
[2]*CAS Center for Excellence in Quantum Information and Quantum Physics, University of Science and Technology of China, Hefei 230026, China*
[3]*Hefei National Laboratory, University of Science and Technology of China, Hefei 230088, China*

**zyzhouphy@ustc.edu.cn*
*†drshi@ustc.edu.cn*

**Abstract:** Given that the phase amplification method based on harmonic generation exhibits significant phase super-resolution capability in interferometric precision measurement, extending this technology to birefringence interferometers to achieve super-resolution characterization of birefringent crystal properties has important research significance and application value. Here, we achieve a four-fold enhancement in the measurement resolution of the thermo-optic coefficient of a $KTiOPO_4$ crystal by combining a self-stabilized birefringence interferometer with cascaded second harmonic generation processes. We observe the tunable interference beating phenomenon by rotating a birefringent crystal versus the temperature of the crystal for the fundamental wave, second harmonic, and fourth harmonic. Furthermore, the fourth harmonic interference fringes beat 4 times faster than the fundamental wave interference fringes. This beating effect is used to determine the thermo-optic coefficients of the two principal refractive axes with a single measurement. This work provides a feasible, real-time, and robust method for super-resolution measurements based on birefringence interferometry.

## 1. Introduction

The thermo-optic coefficient, a critical parameter dictating the temperature-dependent refractive index variation in birefringent crystals like $KTiOPO_4$ (KTP), $LiNbO_3$ (LN), and $BaB_2O_4$ (BBO), plays a pivotal role in optimizing photonic devices, including electro-optic modulators [1], nonlinear frequency conversion [2], and quantum light sources [3]. Accurate measurement of the thermo-optic coefficient is essential for optimizing the design and functionality of these devices under varying thermal conditions. Optical interferometry has been widely used to measure such crystal coefficients due to its non-invasive nature, high precision, and rapid response capabilities [4-6]. However, conventional interferometric methods are inherently constrained by environmental noise, limited phase resolution, and challenges in decoupling subtle thermally induced birefringence changes [7-9]. To address these challenges, the phase amplification method is undoubtedly a powerful tool for achieving super-resolution measurements of phase variations.

One well-known method that is used in quantum optics to realize phase amplification is based on the multiphoton number and path entangled state known as the NOON state [10-12], which utilizes all *N* photons passing through either arm of the interferometer to accelerate the phase oscillation of the interferometer and consequently attaining super-resolution. Subsequently, Zhou et al. utilized a two-photon interferometer to achieve super-resolution measurement of the thermo-optic coefficient of KTP crystal and revealed the interference behavior of photons in a birefringent interferometer [13]. However, it is very difficult to prepare

NOON states with high photon numbers; the highest number of the NOON states to be prepared to date is around 10 [14]. Additionally, the detection probability is very low when $N$ is large, and a high-photon-number NOON state is highly sensitive to any optical losses experienced by the photons [15,16]. Recently, we reported a phase amplification method based on cascaded second harmonic generation (SHG) processes to achieve real-time phase super-resolution measurement, which has stronger robustness than the NOON state-based method [17]. Therefore, achieving super-resolution measurement of the thermo-optic coefficient of birefringent crystals through this phase method is also a subject of significant interest among researchers.

In this work, we first give a theoretical description of a birefringent Mach-Zehnder interferometer (MZI) based on phase amplification and then achieve super-resolution measurement of the thermo-optic coefficient of KTP crystal through a specially designed intrinsically stable birefringent polarization MZI and two polarization-independent SHG modules. After cascaded SHG processes, the oscillation period of the interference curve is reduced to 1/4 of the original, which means that the measurement resolution is improved by 4 times. In addition, the interference beating phenomenon versus crystal temperature is observed for the fundamental wave (FW), second harmonic (SH), and fourth harmonic (FH). The beating intensity can be tuned by rotating the crystal, and the FH interference fringes beat 4 times faster than the FW interference fringes. This beating effect is used to determine the thermo-optic coefficients of the two principal refractive axes with a single measurement. This work overcomes the resolution limitations of traditional interferometry methods and provides a universal framework for characterizing complex thermo-optic behaviors in birefringent crystals.

## 2. Methods

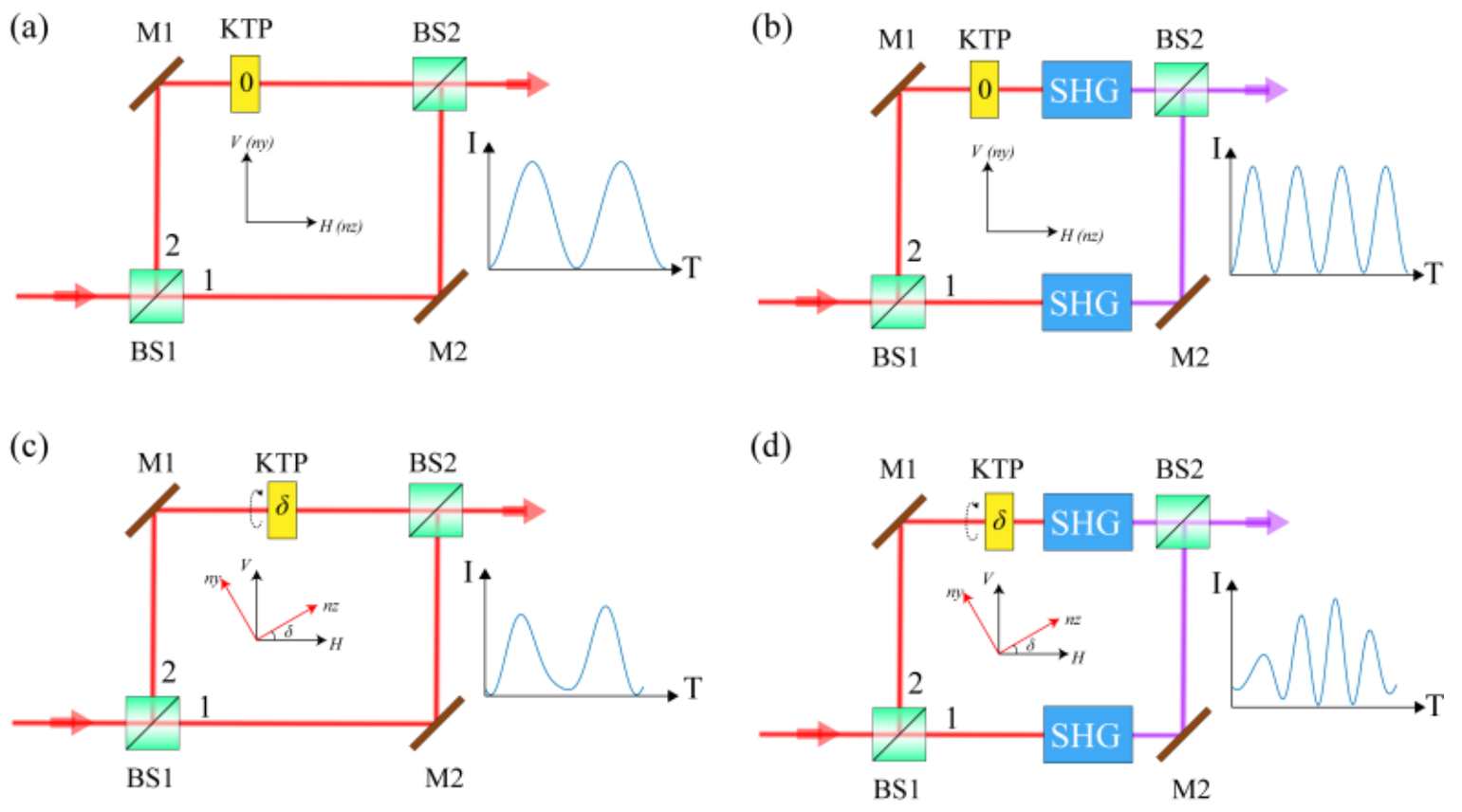

Fig 1. Schematic diagram of super-resolution interferometry principle. (a) Birefringence MZI. (b) Birefringence MZI with KTP crystal rotation angle $\delta$. (c) Nonlinear birefringence MZI including SHG process. (d) Nonlinear birefringence MZI with KTP crystal rotation angle $\delta$.

The general theoretical models will be given first. Figure 1 shows a graphical summary of the main concept of this work. As shown in Fig. 1(a), an x-cut birefringent KTP crystal is inserted into one arm of the MZI, and the y-axis and z-axis of the KTP crystal are in the vertical and horizontal directions, respectively. The vertically polarized FW is split into two beams by the beam splitter (BS). The beam of path 1 passes through the KTP crystal and interferes with the beam of path 2 at the output port of another BS. In wave optics, the light field at the output port of the MZI can be expressed as

$$\tilde{E}_{FW}(\omega)=E_1(\omega)\vec{e}_V+e^{i\Delta\Phi(\omega)}E_2(\omega)\vec{e}_V, \tag{1}$$

where $E_1(\omega)$ and $E_2(\omega)$ represent the two superimposed light fields from path 1 and path 2 at the BS2, $\vec{e}_V$ and $\vec{e}_H$ represent unit vectors in the two orthogonal polarization directions.

$\Delta\Phi(\omega)=2\pi\left[n_y(\lambda_0,T)-1\right]L/\lambda_0$ represents the phase difference between the two superimposed light fields, where $L$ is the length of the KTP crystal and $n_y(\lambda_0,T)$ represents the y-axis refractive index of the KTP crystal related to the fundamental wavelength $\lambda_0$ and temperature $T$. $n_y(\lambda_0,T)$ can be further expanded as $Tdn_y(\lambda_0,T)/dT+n_y(\lambda_0,T_0)$, where $dn_y(\lambda_0,T)/dT$ denotes the thermo-optical coefficient of the y-axis of the KTP crystal. In the following theoretical derivation, it is assumed that $E_1(\omega)=E_2(\omega)$, the interference intensity of the FW can be expressed as

$$I_{FW}(\omega)=2I_1(\omega)\left[1+\cos\left(\frac{2\pi LT}{\lambda_0}\frac{dn_y}{dT}+\varphi_0\right)\right]. \tag{2}$$

Therefore, by continuously tuning the temperature of the KTP crystal, the thermo-optic coefficient of the y-axis can be derived from the period of the resulting cosine interference fringes. As reported in our previous work [17], when we insert SHG modules into the two arms of the MZI, as shown in Fig. 1(b), the oscillation speed of the SH interference fringes will be twice as fast as that of the FW, and the corresponding complex amplitude and intensity of the SH interference light field can be expressed as

$$\begin{aligned}\tilde{E}_{SH}(2\omega)&=E_1(2\omega)\vec{e}_V+e^{i2\Delta\Phi(\omega)+i\Delta\Phi(2\omega)}E_2(2\omega)\vec{e}_V,\\ I_{SH}(2\omega)&=2I_1(2\omega)\left[1+\cos\left(\frac{4\pi LT}{\lambda_0}\frac{dn_y}{dT}+2\varphi_0\right)\right],\end{aligned} \tag{3}$$

where $E_1(2\omega)$ and $E_2(2\omega)$ are the two superimposed SH light fields at the output port of the BS, and the relationships between the SHG beam and the two FW beams are given by $E_1(2\omega)\propto E_1(\omega)\tanh\left(\kappa E_1(\omega)\right)$ and $E_2(2\omega)\propto E_2(\omega)\tanh\left(\kappa E_2(\omega)\right)$, respectively; $\kappa$ is a constant that is proportional to the second-order susceptibility $\chi^{(2)}$ and other experimental parameters of the nonlinear crystal used here. $\Delta\Phi(2\omega)$ is the phase difference generated by the SH beams of the two arms of the interferometer during propagation. From Eqs (1), (2), and (3), we can see that after phase amplification, the equivalent thermo-optic coefficient of the y-axis is increased to twice its original value, which means that the interferometer can more easily distinguish the thermo-optic coefficient of the crystal within a limited temperature tuning range. In addition, this super-resolution capability can be further enhanced by cascading more SHG modules in the two arms of the MZI.

Next, we rotate the KTP crystal in the birefringence MZI around the x-axis (the direction of beam propagation), as shown in Fig. 1(c). The corresponding complex amplitude and intensity of the FW interference light field can be expressed as:

$$\begin{aligned}\tilde{E}_{FW}(\omega,\delta)&=E_1(\omega)\vec{e}_V+E_2(\omega)\left(\alpha\vec{e}_H+\beta\vec{e}_V\right),\\ I_{FW}(\omega,\delta)&=I_1(\omega)\left[1+\sin^4\delta+\cos^4\delta+2\cos^2\delta\sin^2\delta\cos\left(\Delta\varphi_y-\Delta\varphi_z\right)\right.\\ &\quad\left.+2\cos^2\delta\cos\Delta\varphi_y+2\sin^2\delta\cos\Delta\varphi_z\right],\end{aligned} \tag{4}$$

where $|\alpha|^2+|\beta|^2=1$, $\alpha=\cos\delta\sin\delta\left(e^{i\Delta\varphi_z}-e^{i\Delta\varphi_y}\right)$, $\beta=\cos^2\delta e^{i\Delta\varphi_y}+\sin^2\delta e^{i\Delta\varphi_z}=|\beta|e^{i\theta}$, $\delta$ is the rotation angle of the KTP crystal, and $\Delta\varphi_i=2\pi n_i(\lambda_0,T)L/\lambda_0$ $(i=y,z)$ represents the optical phase change of the light field propagating along the y-axis and z-axis of the birefringent crystal. It should be noted that $I_{FW}(\omega,\delta)$ in Eq. (4) has ignored the background noise generated by the horizontal polarization component in $\tilde{E}_{FW}(\omega)$. According to Eq. (4), it can be seen that the FW interference fringes are no longer single-frequency oscillation curves with uniform amplitude but instead exhibit a beat oscillation curve with amplitude modulated periodically. Furthermore, utilizing this beat frequency effect, we can determine the thermo-optic coefficients of the two principal refractive axes with a single measurement. Does this beating phenomenon still exist after phase amplification? If so, is it still possible to achieve super-resolution measurements of the thermo-optical coefficient? Similarly, we rotate the KTP crystal in the nonlinear MZI containing the SHG module by an angle $\delta$, as shown in Fig. 1(d), and the complex amplitude of the interference light field at the BS output port is given by

$$\tilde{E}_{SH}(2\omega,\delta) = E_1(2\omega)\vec{e}_V + e^{i2\theta}E_2(2\omega)\vec{e}_V. \tag{5}$$

The two SH superposition light fields can be given by $E_1(2\omega) \propto E_1(\omega)\tanh(\kappa E_1(\omega))$ and $E_2(2\omega) \propto |\beta| E_2(\omega)\tanh(\kappa|\beta|E_2(\omega))$, respectively. Therefore, the SH interference light field intensity can be expanded as

$$\begin{aligned} I_{SH}(2\omega,\delta) &\propto I_1(\omega)\Big[\tanh^2(\kappa E_1(\omega)) + |\beta|^2\tanh^2(\kappa|\beta|E_2(\omega)) \\ &\quad +2|\beta|\tanh(\kappa E_1(\omega))\tanh(\kappa|\beta|E_2(\omega))\cos 2\theta\Big]; \\ |\beta| &= \sqrt{\sin^4\delta + \cos^4\delta + 2\cos^2\delta\sin^2\delta\cos(\Delta\varphi_y - \Delta\varphi_z)}, \\ \theta &= \tan^{-1}\left(\frac{\sin^2\delta\sin\Delta\varphi_z + \cos^2\delta\sin\Delta\varphi_y}{\sin^2\delta\cos\Delta\varphi_z + \cos^2\delta\cos\Delta\varphi_y}\right); \end{aligned} \tag{6}$$

Comparison of Eq. (4) with Eqs. (5) and (6) demonstrate that the interference curve of the SH also exhibits beating behavior, and the beating curve oscillates twice as fast as that of the FW. Based on the same principle, the complex amplitude of the FH is given by

$$\begin{aligned} &\tilde{E}_{FH}(4\omega,\delta) = E_1(4\omega)\vec{e}_V + e^{i4\theta}E_2(4\omega)\vec{e}_V; \\ &E_1(4\omega) \propto E_1(\omega)\tanh(\kappa E_1(\omega))\tanh(\kappa E_1(\omega)\tanh(\kappa E_1(\omega))), \\ &E_2(4\omega) \propto |\beta|E_2(\omega)\tanh(\kappa|\beta|E_2(\omega))\tanh(\kappa|\beta|E_2(\omega)\tanh(\kappa|\beta|E_2(\omega))). \end{aligned} \tag{7}$$

After the above theoretical derivation, a complete mathematical model of a birefringence interferometer based on the SHG process to achieve phase super-resolution measurement is given. Next, we will further experimentally verify the predictions described in the above theoretical model.

## 3. Experimental setup

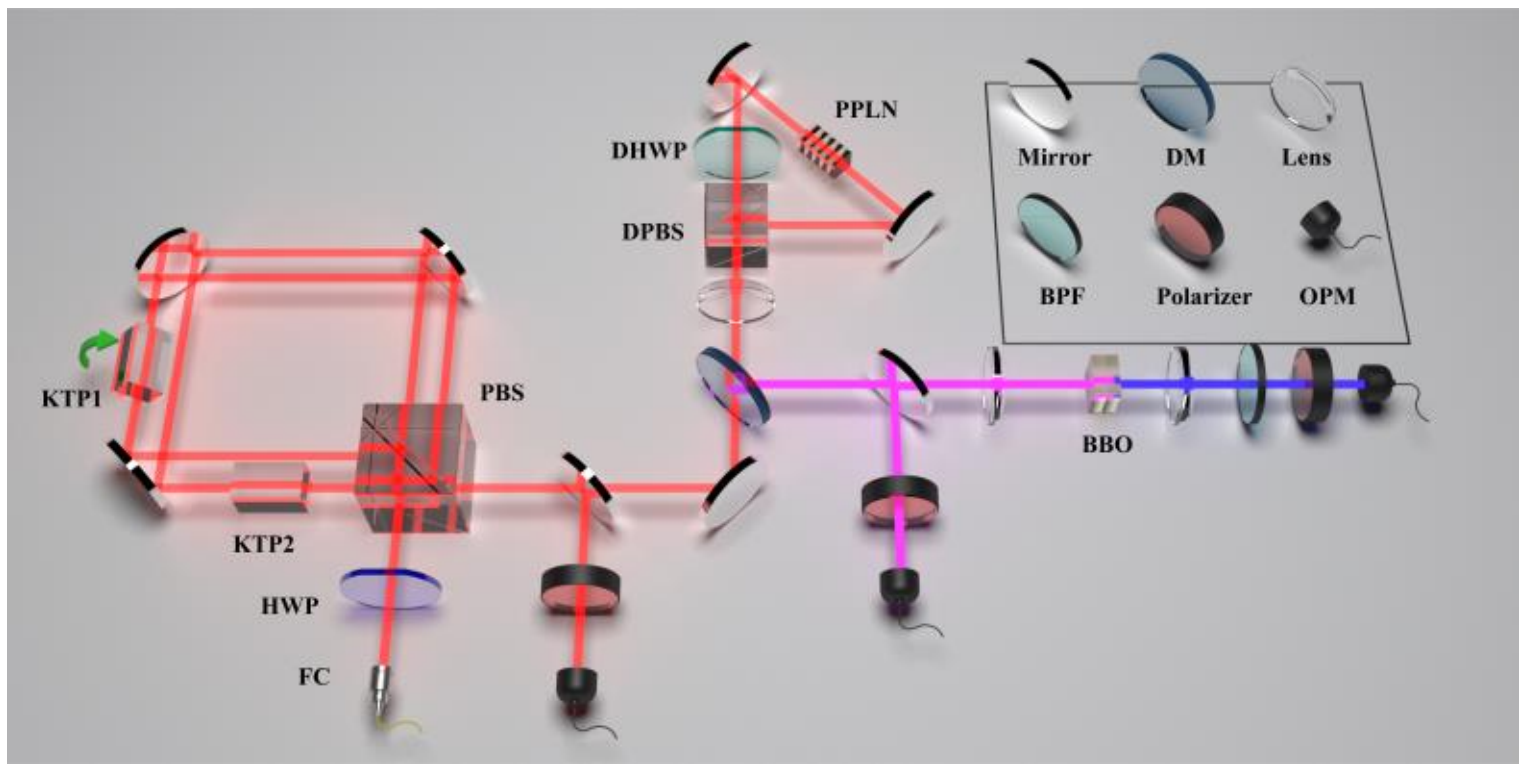

Fig. 2. Experimental setup for super-resolution measurement. FC, fiber collimators; HWP, half-wave plates; DHWP, dichroic HWP; PBS, polarizing beam splitter; DPBS dichroic PBS; KTP, potassium titanyl phosphate crystal; PPLN, periodically poled lithium niobate crystal; BBO, β-barium borate crystal; DM, dichroic mirror; BPF, 390-10 nm band-pass filter; OPM, optical power meter.

A schematic of the experimental setup is shown in Fig. 2. The light source involved in the experiment is a homemade mode-locked fiber laser with a central wavelength of 1560 nm, a pulse duration of 212 ps, and a repetition frequency of 21.6 MHz. After being amplified by a homemade erbium-doped fiber amplifier, the power of the pulsed light can reach 2 W. The linearly polarized pulsed light is first transformed into a 45°-polarized beam using a half-wave plate (HWP), and then injected into a self-stabilized polarization MZI, which contains two KTP crystals; one KTP crystal is used for the measurements, while the other compensates for the optical path length differences between the two arms of the MZI. The self-stable MZI is based on a tilted Sagnac loop, where the clockwise and counterclockwise beams have a traverse distance of 10 mm. Since light beams in the two arms of the MZI are slightly tilted and in counterpropagating configurations, both light beams have nearly the same sensitivity to environmental turbulence, such as temperature fluctuation and vibrations. Both crystals are x-cut so that the beams propagate along the x-axis of the crystal. The two KTP crystals have dimensions of 5 × 5 × 8 mm, and both end faces are antireflection coated for 1560 nm. The temperature of the KTP1 crystal used for the measurement can be tuned from 9 (± 0.002) °C to 32 (± 0.002) °C, while the temperature of KTP2 used for compensation is stabilized at a constant 23 (± 0.002) °C. Moreover, the KTP1 is mounted on a rotation stage to rotate its position with respect to the vertical polarization direction. The orthogonally polarized FW from the two arms of the self-stable MZI enters the first Sagnac-type polarization-independent SHG module, which consists of a dichroic PBS (DPBS), a dichroic half-wave plate (DHWP), and a periodically poled lithium niobate (PPLN) crystal with a length of 25 mm and a poling period of 19.62 μm. The operation temperature of the PPLN crystal is set to 39.4 °C to fulfill the quasi-phase-matching condition. The Sagnac loop with a DHWP inserted is used to realize the SHG for both vertical and horizontal polarizations, which was demonstrated in our previous work [17]. Then, the generated SH is separated from the FW propagation path by a dichroic mirror (DM). The second polarization-independent SHG module is composed of two orthogonally glued β-barium borate (BBO) crystals that satisfy type-I birefringence phase matching conditions. Each BBO crystal has a thickness of 0.5 mm and a phase-matching angle of ~30°. The horizontally and vertically polarized SH can be respectively converted into oppositely polarized FH through the corresponding BBO crystals. Therefore, two orthogonally polarized 1560 nm FW pass through two cascaded SHG modules, successively generating 780 nm SH and 390 nm FH. The 45° polarizer is used to perform projection measurements of two orthogonal polarization states, and the resulting interference fringes are detected and recorded by a power meter. When we need to observe the interference fringes of FW (SH), a mirror is used to separate the FW (SH) beam from the original optical path; otherwise, the mirror is not present in the optical path.

## 4. Results

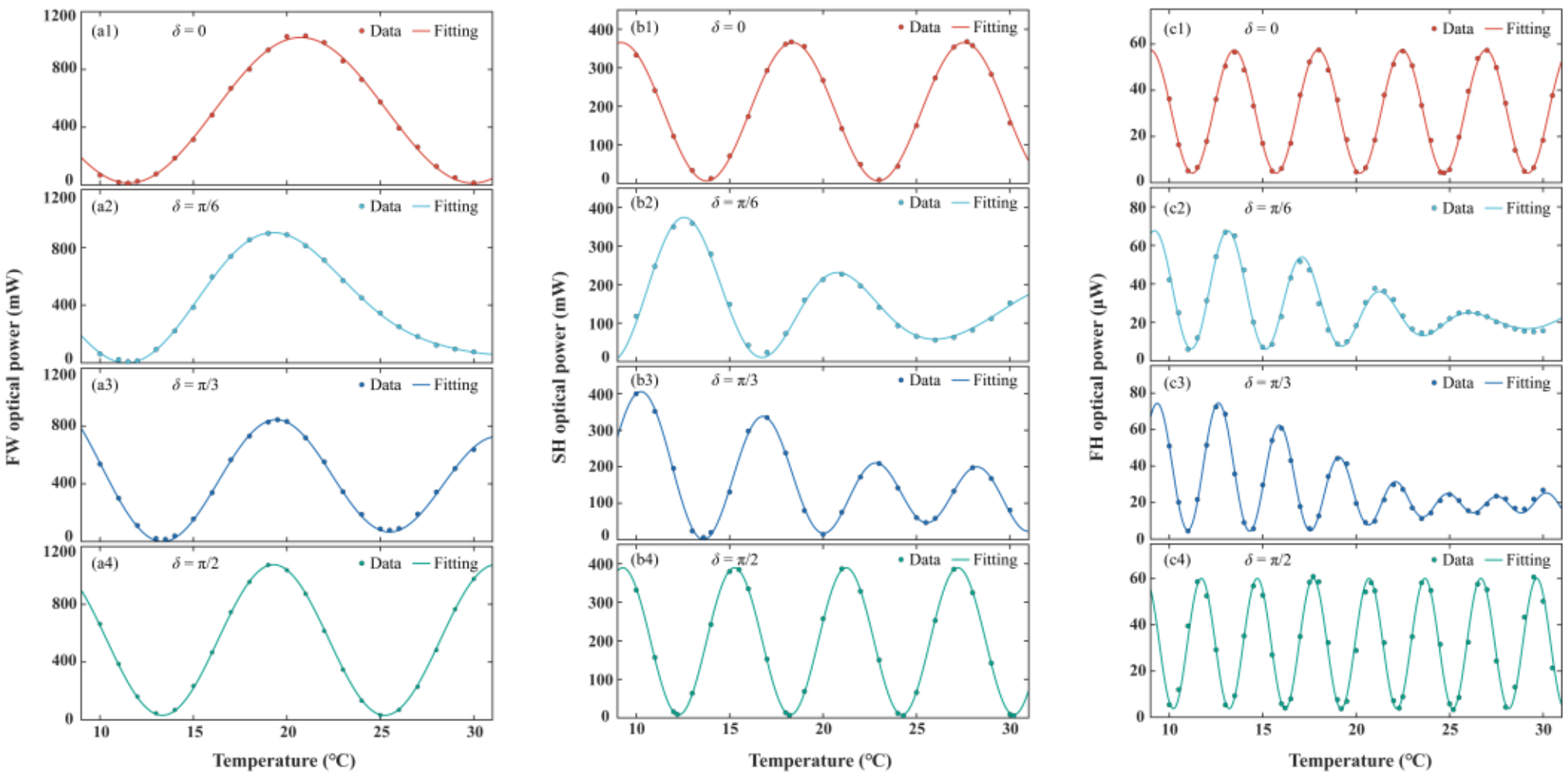

Fig. 3. Interference beating versus temperature for the FW, SH, and FH. The panels on the left (from top to bottom) represent the FW cases at rotation angles of $\delta = 0$, $\pi/6$, $\pi/3$, $\pi/2$. The panels on the middle and right represent the corresponding interference results of the SH and FH, respectively. Different offsets in each of the interference fringes come from the different initial phases between the two arms of the interferometer.

Figure 3 shows the interference fringes of the FW, SH, and FH measured experimentally at different rotation angles of the KTP1 crystal. The x-axis represents the temperature of the KTP crystal, and the y-axis represents the optical power of the measured interference pattern. The panels on the left (from top to bottom) show the beating curves of the FW for rotation angles of $\delta = 0, \pi/6, \pi/3, \pi/2$. The panels on the middle and right represent the corresponding beating curves of the SH and FH, respectively. The solid lines are the fitting curves obtained from the theoretical model described above. The rotation angles of $\delta = 0, \pi/2$ represent cases in which the input light polarization coincides with the y and z optical axes of the birefringent crystal. In these two cases, the periods of the interference curves of the SH and FH are reduced to 1/2 and 1/4 of that of the FW interference curve, respectively, which indicates that higher measurement resolution is achieved after phase amplification. The interference visibility of the interference fringes for the FW, SH, and FH at $\delta = 0$ ($\pi/2$) is 97.4% (94.3%), 96.6% (96%), and 86.4% (88.3%), respectively. The relatively low interference visibility of the FH is primarily attributed to the elongation of the FH beam spot caused by the walk-off effect during the SHG process in the BBO crystal. This results in incomplete spatial overlap between the horizontally and vertically polarized FH fields. Consequently, the non-overlapping regions of the light fields contribute to background noise during interference, thereby reducing the visibility of the interference fringes. Based on the interference fringes of the FW, SH, and FH under the condition of $\delta = 0$ ($\pi/2$), the thermo-optic coefficients of the KTP crystal along the y (z) axis were determined to be $1.062 \times 10^{-5}/K$ $\left(1.639 \times 10^{-5}/K\right)$, $1.062 \times 10^{-5}/K$ $\left(1.631 \times 10^{-5}/K\right)$, and $1.086 \times 10^{-5}/K$ $\left(1.630 \times 10^{-5}/K\right)$, respectively.

When the polarization direction of the FW does not coincide with the principal axis of the birefringent crystal, the interference fringes exhibit beating behavior of the optical properties along the two axes, and we can determine the optical properties along both axes from any single measurement of this type of beating curve. For example, when $\delta = \pi/3$, the thermo-optic coefficients of the y and z axes obtained from the beating curves of FW (SH, FH) were $0.980 \times 10^{-5}/K$ $\left(0.955 \times 10^{-5}/K, 1.023 \times 10^{-5}/K\right)$ and $1.593 \times 10^{-5}/K$ $\left(1.606 \times 10^{-5}/K, 1.618 \times 10^{-5}/K\right)$, respectively. Another essential feature of the beating curve is that the temperature oscillation periods of the beating curve for the SH and FH are 2 times and 4 times faster than that of the FW, respectively, which demonstrates that super-resolution measurement of the thermo-optic coefficients of birefringent crystals can still be achieved under beating conditions. In addition, as the phase amplification factor increases, a higher measurement resolution can be achieved. It should be noted that although the actual environmental temperature fluctuation exceeds 0.002 °C (usually around ± 0.5 °C), the relative phase of the interferometer remains stable throughout the measurement period. This indicates that the self-stabilized polarization MZI exhibits excellent resistance to environmental turbulence.

Next, we characterize the conversion efficiency of the two polarization-independent SHG modules. In our experiment, the polarization-independent SHG modules exhibit nearly identical conversion efficiency characteristics for both horizontally and vertically polarized input light. Therefore, we show the power conversion efficiency of the two SHG modules in the case of vertically polarized input. The results are shown in Fig. 4, where the dashed lines are the theoretical fitting curves. The nonlinear efficiency of the first SHG process in the experiment exceeds 30%, while that of the second SHG process is less than ~ 0.1%. This is due to two main factors: first, the pump light intensity entering the second SHG process is significantly reduced after the first SHG process; second, the nonlinear crystal used in the second SHG process is 0.5 mm thick BBO, which has a much smaller nonlinear coefficient compared to PPLN. In addition, for the SHG process based on the BBO crystal, the walk-off effect also leads to a decrease in conversion efficiency. As presented in our previous work, under optimal experimental conditions (with a high-intensity pump laser and a proper crystal), the power conversion efficiency of SHG can be improved to 80% [18]; therefore, much higher levels of super-resolution measurements can be achieved in principle.

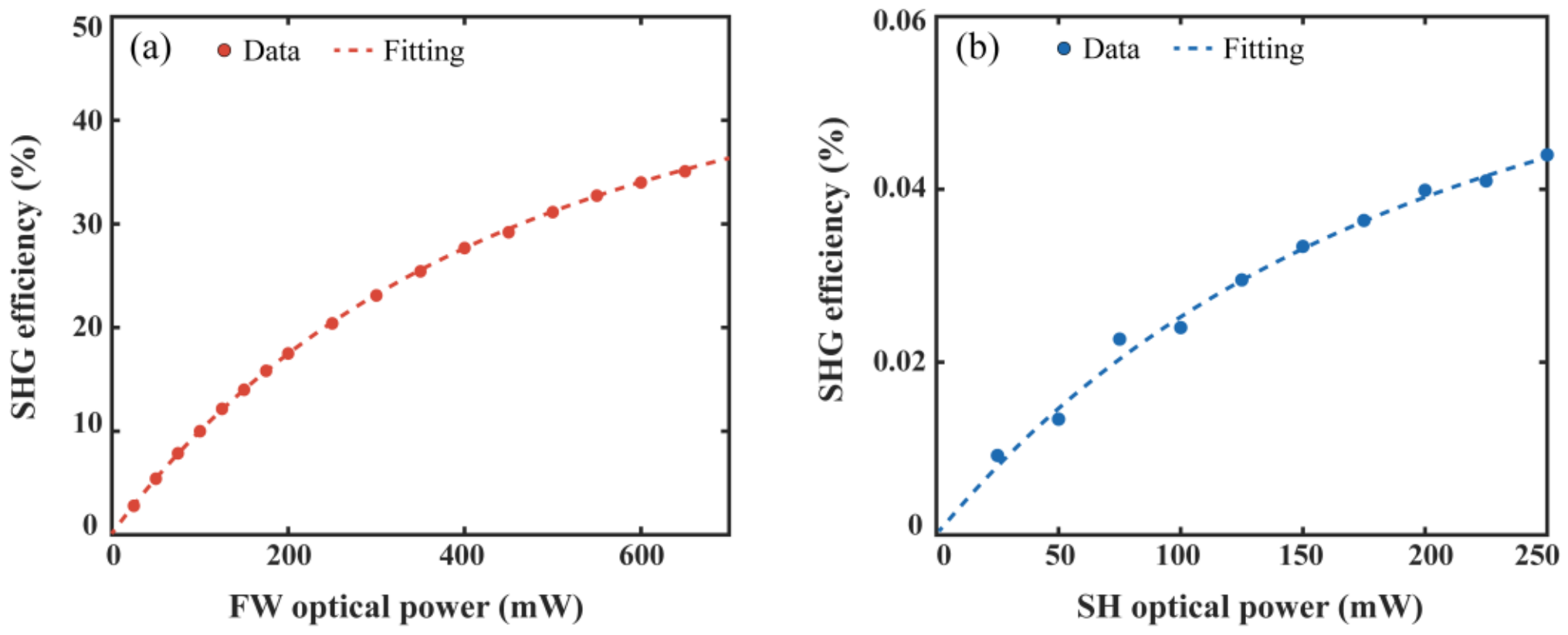

Fig. 4. Power conversion efficiency of the (a) first and (b) second polarization-independent SHG modules.

## 5. Conclusions

In summary, we have achieved super-resolution measurement of the thermo-optic coefficient of KTP crystal using birefringence interferometry based on the phase amplification method. Through the carefully designed self-stabilized birefringent MZI and polarization-independent SHG modules, the physical mechanism of super-resolution interferometric measurements is revealed in detail. After phase amplification, the FH interference fringes oscillate four times faster than that of the FW, which indicates a fourfold enhancement in resolution capability. In addition, the interference beating phenomenon versus the crystal temperature has been observed for the FW, SH, and FH. This beating feature is used to determine the optical properties along both crystal axes with a single measurement. Although we have only achieved a fourfold improvement in resolution here, as described in our previous work [17], further optimization of the pump light source and nonlinear crystal parameters, along with the application of optical parametric amplification techniques, can enable even higher measurement resolution. We should point out that the present system is not limited to the determination of the thermo-optic coefficient of a birefringent crystal and can also be used to determine the wavelength dispersion [19] and the electro-optical coefficient of the birefringent crystal. More importantly, compared to methods for achieving super-resolution measurements based on the NOON state, our scheme boasts real-time responsiveness and is compatible with existing mature interferometric measurement systems. This work will thus be of great importance for understanding super-resolution measurements based on phase amplification.

**Funding.** National Key Research and Development Program of China (2022YFB3903102, 2022YFB3607700), National Natural Science Foundation of China (NSFC) (62435018), Innovation Program for Quantum Science and Technology (2021ZD0301100), USTC Research Funds of the Double First-Class Initiative, and Research Cooperation Fund of SAST, CASC (SAST2022-075).

**Acknowledgment.** We sincerely thank Yinhai Li, He Zhang, and Mengyu Xie for their help in this work.

**Author contributions.** Z.-Y.Z. and W.-Z.L. designed the experiment. W.-Z.L. carried out the experiment. W.-Z.L. analyzed the data and wrote the paper with input from all other authors. The project was supervised by Z.-Y.Z. and B.-S.S. All authors discussed the experimental procedures and results.

**Disclosures.** The authors declare no conflicts of interest.

**Data availability.** Data underlying the results presented in this paper are not publicly available at this time but may be obtained from the authors upon reasonable request.